# Chapter 6
# Judicial Review of Uncertain Risks in Scientific Research

**Eric E. Johnson**

**Abstract** It is difficult to neutrally evaluate the risks posed by large-scale leading-edge science experiments. Traditional risk assessment is problematic in this context for multiple reasons. Also, such experiments can be insulated from challenge by manipulating how questions of risk are framed. Yet courts can and must evaluate these risks. In this chapter, I suggest modes of qualitative reasoning to facilitate such evaluation.

## 6.1 Introduction

In the world of engineering, uncertainty is anathema. Whether designing a nuclear power plant or planning a system of levees, engineers do not like surprises. Experimental scientific research, however, thrives on uncertainty. Reaching beyond the current state of knowledge is the whole point of the enterprise. Whether probing the planets or trying to create new particles, scientists would love to see a surprise around every corner. Uncertainty goes hand in hand with aspirations to explore and discover.

This special role for uncertainty in scientific research makes questions of catastrophic risk from science experiments particularly interesting. What is more,

---

[1]See Hawai'i County Green Party v. Clinton, 980 F. Supp. 1160, 1168-69 (D. Haw. 1997); Karl Grossman, The Risk of Cassini Probe Plutonium, The Christian Science Monitor (Oct. 10, 1997) http://www.csmonitor.com/1997/1010/101097.opin.opin.1.html; CNN, Cassini roars into space, CNN (Oct. 15, 1997) http://www.cnn.com/TECH/9710/15/cassini.launch/; Najmedin Meshkati, Probability, Plutonium Don't Mix (op-ed), Los Angeles Times (Oct. 10, 2013) http://articles.latimes.com/print/1997/oct/10/local/me-41149.

E.E. Johnson (✉)
University of North Dakota, Grand Forks, USA
e-mail: eric.e.johnson@law.und.edu

science experiments have raised the specter of catastrophe scenarios noteworthy for their exoticness and magnitude. Critics have alleged that a plutonium-laden interplanetary spacecraft may pose the risk of causing millions of cancers by disintegrating in the atmosphere.[1] A microbiological laboratory could plausibly cause a pandemic with an accidental release of an exotic pathogen [1]. And experimenting at the edge of knowledge of fundamental physics could, in the view of some, risk destroying the entire planet [11].

In this chapter, I will look at uncertainty in science-experiment risk from a legal perspective. Suppose someone brings an action in court to halt an experimental endeavor based on the allegation that it poses an undue risk to the public. Courts generally have the power to order a halt to activity that puts human life at risk.[2] But how do courts decide that there is a substantial chance of human casualties? It's not an easy task under the best of circumstances. And when the context is an activity intentionally designed to reach beyond the current state of human knowledge, it is especially tricky.

Courts have often side-stepped thorny issues of uncertainty, embracing instead the traditional view of risk.[3] In the traditional view, risk can be understood in terms of probabilities and applied in the form of cost-benefit balancing [22]. Traditional risk evaluation, then, depends heavily on experts and quantitative analysis.[4] This mode of thinking corresponds to classical theories of decision-making, in which information is available, problems are known, and optimal solutions can be found through the application of perfect reason [10, p. 123]. This traditional paradigm is broadly problematic because it ignores the existence of human limitations in understanding. In other words, it ignores the central role of uncertainty in risk.

---

[2]In common-law jurisdictions, this may be accomplished through the court's inherent equitable power. See, e.g., Harris Stanley Coal & Land Co. v. Chesapeake & Ohio Railway Co., 154 F.2d. 450 (6th Cir. 1946) (reversing district court's refusal to enjoin coal mining operations that posed a risk of subsidence to railroad tracks, because "[a] court of equity will not gamble with human life, at whatever odds, and for loss of life there is no remedy that in an equitable sense is adequate."); Shimp v. New Jersey Bell Telephone Co., 368 A.2d 408 (N.J. Super. Ch. 1976) (recognizing a "common-law right to a safe working environment" and finding that "such work area was unsafe due to a preventable hazard which the court had power to enjoin."). Courts in civil-law jurisdictions have similar powers conferred through broad provisions of the civil law.

[3]To say that courts often embrace a very traditional view of risk and side-step uncertainty is not to say they do so always. Particularly in the context of reviewing administrative regulations, courts have sometimes embraced the precautionary principle, the essence of which is to err on the side of caution in conditions of uncertainty. See, e.g., Alberto Alemanno, The Shaping of the Precautionary Principle by European Courts—From Scientific Uncertainty to Legal Certainty (Bocconi Legal Studies Research Paper No. 1007404, Aug. 16, 2007) (discussing the role played by European Community courts in shaping the precautionary principle); Indus. Union Dept., AFL-CIO v. Am. Petroleum Inst., 448 U.S. 607, 656 (U.S. 1980) ("the Agency is free to use conservative assumptions in interpreting the data with respect to carcinogens, risking error on the side of overprotection rather than underprotection").

[4][22, p. 123]. See also Chap. 4.

In many cases coming before the courts—such as garden-variety products-liability actions—the classical approach might be relatively unobjectionable. Even though uncertainty exists in all situations of risk, perhaps it is often efficient for courts to disregard uncertainty and proceed as if risk is necessarily tamable by number-crunching experts.

But when it comes to science-experiment catastrophe risk, the role of uncertainty cannot be ignored. The traditional paradigm for thinking about risk is especially problematic in this context for at least three reasons. First, reliance on experts is particularly troublesome since the only persons who can supply expert opinions on the safety of leading-edge science experiments are often the same people who are involved in the experiment, either directly or indirectly. Second, working in the certainty mode of risk assessment may require the application of scientific knowledge that the experiment itself is designed to discover, creating a kind of catch-22. Third, where the computed probability of science-experiment disaster is small, it is overwhelmed by uncertainties in the assessment. That is, the chance the probability assessment is wrong—for instance, because of erroneous assumptions, flawed calculations, or observational errors—is many times greater than the computed probability itself.

Beyond these practical and conceptual concerns, there is also a problem of rhetoric. Traditional risk assessment can be easily frustrated and manipulated by science-experiment proponents. In leading-edge science, the close-knit community of researchers championing the experiment may have an effective monopoly on expertise and knowledge in the area, and this can allow those scientists to control the framing of risk questions. This framing can be used to frustrate meaningful engagement on safety issues in terms of traditional risk analysis. And selective assertions of uncertainty can steer the debate toward a favorable outcome for experiment supporters.

I will explore these concepts primarily through one fascinating and extreme example—CERN's Large Hadron Collider (LHC), which critics say could create a planet-destroying black hole.[5] After pointing out the various troubling features of science-experiment risk, I propose a way for the courts to deal meaningfully with legal challenges to such experiments despite the many difficulties.

## 6.2 Background on the LHC/Black-Hole Disaster Scenario

In this part, I will provide some background on the LHC, the laboratory organization that operates it, and the black-hole risk the experiment is said to pose.

---

[5]The instant paper follows on a previous article I wrote that discusses some aspects of the black-hole issue in additional detail and analyzes some other issues, such as jurisdiction, evidence, injunction law, and jurisprudence [11].

### *6.2.1 CERN and the LHC*

Located near Geneva, CERN is the world's pre-eminent laboratory for particle physics, which studies the most fundamental aspects of matter. An intergovernmental organization comprising 21 member states, CERN is a mammoth institution.[6] Its 2014 budget was 1.1 billion CHF (€1.0 billion).[7] And a recent count indicated that 10,357 experimental physicists were involved in CERN in some capacity.[8]

CERN's current program is centered around the LHC, a superconducting synchrotron particle accelerator, along with its various component experiments.[9] Launched in 2008, the LHC is, according to CERN, "the largest machine in the world".[10] The beam tunnel is 27 km around.[11] The magnets used to accelerate the particles are cooled with about 90 tons of superfluid helium. And as a whole, the apparatus consumes enough power for a medium-sized city.[12]

The aim of the LHC is to advance understanding of the fundamental particles and forces that make up the physical universe.[13] Indeed, the LHC already achieved a triumphant discovery by finding the celebrated Higgs boson—a fundamental particle that is understood to impart mass to matter.[14]

The LHC was designed to collide protons at an energy of 14 trillion electron volts (TeV), reaching a new energy range that scientists have dubbed the "terascale".[15] This collision energy will be 14 times the energy of any previous experiment.[16] As of the spring of 2015, the LHC was just beginning to get close to achieving this energy, coming out of a two-year "long shutdown" involving intensive maintenance.[17] During its first run, the LHC had been limited to a collision energy of 8 TeV.[18]

---

[6] See CERN, CERN brochure 1 (2010), http://cds.cern.ch/record/1278456/files/CERN-Brochure-2010-005-Eng.pdf.

[7] See CERN General Information 2014, http://cds.cern.ch/record/1745540/files/CERN-Brochure-2014-006-Eng.pdf.

[8] See CERN General Information 2012, http://cds.cern.ch/record/1453354/files/CERN-Brochure-2012-003-Eng.pdf.

[9] See CERN, CERN FAQ LHC: The Guide 4, 15-19 (2008), available at http://cdsmedia.cern.ch/img/CERN-Brochure-2008-001-Eng.pdf.

[10] CERN, All timelines | CERN timelines, http://timeline.web.cern.ch/timelines.

[11] CERN, CERN FAQ LHC The Guide at 19, http://cds.cern.ch/record/1165534/files/CERN-Brochure-2009-003-Eng.pdf.

[12] Travis Lupick, B.C. Scientists Aim to Unlock Secrets of Universe, The Georgia Straight, May 15, 2008, http://www.straight.com/article-145556/end-world.

[13] See CERN, Why the LHC, http://public.web.cern.ch/public/en/LHC/WhyLHC-en.html.

[14] See Jacob Aron, Elusive Higgs wins physics Nobel, shared with Englert, New Scientist (Oct. 8, 2013), http://www.newscientist.com/article/dn24365-elusive-higgs-wins-physics-nobel-shared-with-englert.html.

[15] CERN, CERN faq LHC The Guide at 3, http://cds.cern.ch/record/1165534/files/CERN-Brochure-2009-003-Eng.pdf; Particle Physics Project Prioritization Panel, US Particle Physics: Scientific Opportunities, A Strategic Plan for the Next Ten Years 2 (2008), http://www.er.doe.gov/hep/files/pdfs/P5_Report%2006022008.pdf.

[16] See Fermilab, Fermilab's Tevatron, http://www.fnal.gov/pub/science/accelerator/.

Early on, the LHC suffered a couple of unanticipated mishaps that caused parts of the machine to blow up.

In 2007, during a test, a design defect caused one of the magnet units to explode. In explaining the error that led to the accident, the director of Fermilab, the American particle physics laboratory that built the magnet, said, "[W]e are dumbfounded that we missed some very simple balance of forces. Not only was it missed in the engineering design but also in the four engineering reviews carried out between 1998 and 2002 before launching the construction of the magnets".[19]

Then in 2008, shortly after the LHC's launch, a faulty electrical connection caused a mishap that damaged 53 of the LHC's magnet units.[20] Although CERN's initial reports characterized the event as a "leak", Cal Tech physicist Sean Carroll offered that "'explosion' is a more accurate description".[21] Magnets were ripped out of their floor bolts and six tons of helium spewed into the tunnel in a matter of minutes.

In 2015, CERN finished two years of refitting work to allow the LHC to operate at 13 TeV, much closer to the machine's design capacity.[22]

As an effort to explore unknown realms of physics, the excitement revolves around uncertainty. In the words of Dan Tovey, a University of Sheffield physicist on the LHC's Atlas research team:

> Individually, we all have the things that we're particularly interested in; there's a variety of new physics models that could show up. But to be honest, we can't say for certain what—if anything—will show up. And the best thing that could possibly happen is that we find something that nobody has predicted at all. Something completely new and unexpected, which would set off a fresh programme of research for years to come.[23]

---

[17]Caroline Duc, Long Shutdown 1: Exciting times ahead, CERN Updates, http://home.web.cern.ch/about/updates/2013/02/long-shutdown-1-exciting-times-ahead (Feb. 8, 2013, updated Dec. 11, 2014); Discovery News, LHC Shuts Down (Temporarily), http://news.discovery.com/space/lhc-shutdown-cern-higgs-130214.htm; Jonathon Webb, LHC Smashes Energy Record with Test Collisions, BBC News (May 21, 2015), http://www.bbc.com/news/science-environment-32809636.

[18]CERN, Press Release: LHC physics data taking gets underway at new record collision energy of 8TeV (Apr. 5, 2012), http://press.web.cern.ch/press-releases/2012/04/lhc-physics-data-taking-gets-underway-new-record-collision-energy-8tev.

[19]Pier Oddone, Directors Corner: The World Stage, Fermilab Today (Apr. 3, 2007), http://www.fnal.gov/pub/today/archive/archive_2007/today07-04-03.html.

[20]Large Hadron Collider Ready to Restart, Boston.com (Nov. 20, 2009), http://www.boston.com/bigpicture/2009/11/large_hadron_collider_ready_to.html.

[21]Sean Carroll, The Particle at the End of the Universe: How the Hunt for the Higgs Boson Leads us to the Edge of a New World location 1100 of 5498 (Dutton 2012).

[22]Webb, supra note 17 (quoting physicist Dan Tovey).

[23]Id. (quoting Tovey; internal paragraph break omitted).

## *6.2.2 Black Holes and the Evolving Safety Rationale*

The alleged risk of the LHC is that it could spawn a planet-destroying black hole. Explaining the thinking about an artificial black-hole disaster is best done in a chronological manner. That is because the safety argument has not been static; instead, it has changed over time. As one set of safety rationales has been undermined by evolving understandings in theoretical physics, experimenters have abandoned old arguments and adopted new ones.

Questions circulating in the media about whether particle colliders might produce black holes date back at least to 1999, around the time of the start-up of the an earlier experiment—the Relativistic Heavy Ion Collider (RHIC, pronounced "Rick"), located about 100 km east of New York City [12, p. 542–553].[24]

The dust-up about the RHIC led physicists in 1999 to issue an assurance that, for the foreseeable future, no particle collider would be capable of generating the energies required to form a black hole [3]. Media interest in the question then subsided. But it turned out that the physicists' safety pronouncements in 1999 were substantially mistaken. In 2001 two separate teams of theorists demonstrated that, under certain assumptions, it actually was possible to produce black holes with a present-day accelerator—the LHC, which was then under construction [4, 8].

CERN subsequently acknowledged a need for further safety assessment work [2]. The ensuing report, issued by a CERN group in 2003, concluded that accelerator-produced black holes would pose no threat since they would rapidly evaporate via a process called "Hawking radiation" [11, p. 840 et seq.].

The next year, 2004, a very well regarded scientist called the theory of Hawking radiation into question [20]. Since the scientist was a pioneer of black-hole-evaporation theory, his opinion was potentially influential.

CERN then stopped relying on Hawking radiation as a safety rationale, and the organization commissioned a new round of theoretical work on the issue [11, p. 850]. The result was a highly complex paper, released in 2008, that rested an assurance of safety on a multi-faceted approach [7]. Authored by particle physicists Michelangelo L. Mangano and Steven B. Giddings, the paper used astrophysics analysis to conclude that, under some scenarios, synthetic black holes would be able to harmlessly coexist with Earth, since they would grow too slowly to be dangerous. Under other scenarios, the paper concluded, telescope observations of certain white dwarf stars could be counted upon to rule out the dangers on an empirical basis.

Rainer Plaga, an astrophysicist, then emerged with a paper arguing that the black hole scenario could not be ruled out [15]. In the days leading up to the anticipated start up of LHC collisions, Mangano and Giddings responded to some, but not all, of Plaga's arguments [6].

---

[24] The controversy over the RHIC principally involves not black holes, but a disaster scenario involving the creation of a "strangelet," which theoretically could physically collapse the planet into a small hyperdense ball by converting all normal matter on Earth into strange matter. I discuss the RHIC more in my prior work [11, p. 829–831] [12, p. 542–553].

The chronology of the safety debate, with its pattern of confident conclusions, new revelations, and refortification with different theoretical tacks, could be seen as evidence of results-oriented research lacking academic detachedness. Indeed, in 2010, John Ellis, a top theoretical physicist for CERN who worked on a safety report seemed to confirm this. Ellis told Physics World magazine that there had been no scientific motivation for the safety reviews, calling them a "foregone conclusion".[25]

None of this, of course, means that the LHC itself is necessarily dangerous. But it does indicate that the question of the LHC's safety is more complex than it might seem at first glance.

## 6.3 Conceptual and Practical Problems

The LHC/black-hole question can be used to illustrate a number of problems that may occur with traditional risk assessment in the context of leading-edge scientific experimentation. To recapitulate from the introduction, they are (1) that it is difficult to rely on experts for conducting such analysis, since those experts tend to have personal connections to the experimental work being analyzed, (2) the risk assessment may require the application of uncertain science that the experiment itself is designed to illuminate, and (3) if the probability of disaster is calculated to be very low, then that probability number is rendered virtually meaningless, since the probability of error in the derivation dwarfs the derived disaster probability.

### 6.3.1 The Lack of Disinterested Experts

Traditional risk analysis is "distinctly expert-centered" [22, p. 123]. But when it comes to science-experiment risk, there may be a scarcity of disinterested experts. With leading-edge science experiments, the leading experts tend to be the exact same people who are involved in the experiment—either directly or indirectly.

The LHC/black-holes question is a case in point. According to CERN, "half the world's particle physicists" come to CERN for their research.[26] The other half are an extended network of friends and acquaintances. Sharon Traweek, an anthropologist who did a particle-physics ethnography, describes particle physicists as forming a restrictive, cohesive community [19]. Relationships among particle physicists are highly important, and she explains that those particle physicists who do not know each other well, want to. Thus, if one wanted to find particle physicists not part of the broader circle of friends of an allegedly dangerous experiment, doing so might prove impossible.

---

[25] See Edwin Cartlidge, Law and the end of the world, Physics World Feb. 2010 at 12–13 (quoting Ellis: "'Every time someone comes up with a new theoretical speculation about accelerator safety, it is interesting to see why that speculation does not constitute risk, but it always comes back to the cosmic-ray argument', he says. So does that mean these safety reviews are nothing more than a curiosity? 'Correct. There is no scientific motivation for these reviews. They are a foregone (Footnote 25 continued)
conclusion, even though the community has the right to expect CERN to demonstrate the validity of the safety arguments.'").

[26] See CERN, A Global Endeavour, http://public.web.cern.ch/public/en/About/Global-en.html; see also Elizabeth Kolbert, Crash Course, The New Yorker, May 14, 2007, at 68 ("Once the collider begins operating at full power [...] nearly half the particle physicists in the world will be involved in analyzing its four-million-megabyte-per-hour stream of data.").

In the case of the LHC, however, the primary risk assessment work was not done by people with mere indirect relations. Instead, the work has been done by persons employed by or having direct ties to CERN. In 2007, CERN management set up the LHC Safety Assessment Group (LSAG),[27] and each of the five members were physicists from CERN's Theory Division [5]. One of those, Mangano, co-authored the paper that served as the foundation for LSAG's final report [7, p. 1]. The other author of that key paper, Giddings, was not employed by CERN while he was working on the paper, but he was at that time anticipating a visiting position with CERN that had been previously approved [11, p. 839].

### 6.3.2 The Need for Uncertain Scientific Principles Under Investigation

A second problem that may occur in trying to resolve questions of the safety of novel science experiments is that a thorough traditional risk assessment might require knowledge that the experiment itself is designed to supply.

Again, the LHC/black-holes question provides examples. The theorized phenomenon of Hawking radiation—which has been used as a safety rationale for particle collisions—has not been experimentally validated [17, p. 172]. The LHC experiment itself, however, could verify Hawking radiation.

Another unanswered question from physics, which the LHC could help answer,[28] is whether there are hidden dimensions, and if so, how many exist. The LHC may create black holes, the theory goes, only if there are one or more extra dimensions [8]. The number of extra dimensions is important to the analysis, because the issue of how fast a stable black hole could grow inside the Earth is understood to depend on the number of hidden dimensions. The Giddings and Mangano paper estimated, for instance, that if there are nine or more spacetime dimensions, it would take billions of years for any black hole to grow large enough to be threatening [7]. But if we live in a 5-D reality—that is, with just one extra dimension—then under Giddings and Mangano's analysis it might only take 300,000 years until the black hole matured and ate the Earth. Again, the ultimate level of risk of black holes rests on unknown science that the LHC may or may not shed light on.

---

[27]Video recording: John Ellis, CERN Colloquium: The LHC Is Safe (Aug. 14, 2008), available at http://cdsweb.cern.ch/record/1120625 (beginning at 6 min.).

[28]See CERN, Extra Dimensions, http://press.web.cern.ch/backgrounders/extra-dimensions.

### 6.3.3 The Effect of Uncertainty in Low-Probability Assessments

The third problem with traditional risk assessment for leading-edge science experiments is that where the probability of disaster is determined to be low, then the probability number lacks robustness, since the chance of disaster described by the probability will be much lower than the chance that the probability itself is wrong [11, p. 883 et seq].

Physicists have not published any quantification of the odds of a black-hole disaster at the LHC. In spoken remarks at Oxford's Future of Humanity Institute, however, Mangano spoke[29] of probabilities of less than $10^{-40}$.

That is indeed a small number. Even when the alleged harm is the annihilation of Earth, it seems perfectly safe to ignore such an infinitesimal chance. But as particle physicist Lisa Randall said, albeit in a different context, "A prediction of low risk is meaningless if the uncertainties associated with the underlying assumptions are much greater" [17, p. 181].

Philosophers Toby Ord, Rafaela Hillerbrand, and Anders Sandberg put it this way: "When an expert provides a calculation of the probability of an outcome, they are really providing the probability of the outcome occurring, given that their argument is watertight" [14, p. 1]. Reasons the argument could be defective include mathematical miscalculations, faulty data, and uncertain assumptions.

Taking, for argument's sake, that a given probability bound is low enough that it represents insignificance, the question is left begging to what extent the estimate itself is flawed. We cannot directly measure the uncertainty of the LHC safety assessment work, as doing so would require having unavailable knowledge. We can, however, borrow error rates that have been empirically determined for other forms of scientific work. One study found that as many as one in 100 articles in the life sciences may contain errors warranting retraction [14, p. 4]. Other research found that one in 10 articles in Nature and the British Medical Journal—both elite journals—have flawed statistical results [14, p. 7]. Using these statistics as a gauge of how likely it is that the LHC risk assessment is flawed, we must say that a truer view of the maximum probability of disaster at the LHC must take into account something like a one-in-10 or one-in-100 chance that the safety assessment is in error. Of course, if the safety assessment were wrong, that would not imply that the LHC is necessarily dangerous. Nonetheless, the total ceiling on risk of operating the LHC must mostly be described by the chance that the safety assessment is flawed. And that probability, even if small, is plausibly significant.

---

[29]See Video Recording: Global Catastrophic Risks Conference, Expected and Unexpected in the Exploration of the Fundamental Laws of Nature (Michelangelo L. Mangano 2008), available at http://vimeo.com/4704040 (address at University of Oxford Future of Humanity Institute) (beginning at 47 min.) (stating that the probability that a particular white dwarf of certain characteristics would have persisted despite the laws of physics being such that they would allow dangerous black-hole formation to take place at the LHC is less than $10^{-40}$, and further noting that this would be the probability for the survival of one such star, and there are multiple such stars).

## 6.4 Rhetorical Issues

In addition to the above described problems, there is an additional wrinkle for risk assessment in the science-experiment context. Knowledge asymmetries and opportunities for selective assertions of uncertainty can be used to reframe questions of risk. This reframing can permit experiment proponents to gain a substantial rhetorical advantage, steering public discourse in a way that is helpful for experimenters.

At the heart of this issue is something of a puzzle: Considering that traditional risk analysis involves cost-benefit scrutiny and relies on calculations and statistics [22, p. 123], one might think that this mathematically based way of evaluating risk would be embraced by scientists. Yet this is not necessarily the case. In a twist, when it came to the black-hole issue, particle physicists moved away from expressing risk in quantitative terms and instead moved to speaking of risk in qualitative terms.

### 6.4.1 Using Pricelessness to Avoid Quantitative Analysis of Benefits

In producing quantified probabilities of harm, traditional risk assessment thrives when it is paired with quantified benefits to use as a point of comparison. Thus, one way proponents of science experiments can avoid a negative assessment in the traditional-risk-analysis mode is to frustrate the quantification of the benefits of the experiment, thus making it incomprehensible to a cost-benefit formula.

Proponents of particle experiments will generally concede that, viewed as an investment, particle physics generates "no return".[30] At first glance, this appears to be an admission against self-interest. But, as American judge and legal scholar Richard Posner pointed out in discussing RHIC safety issues, this seeming weakness is actually a source of strength: "[I]t stumps people who people who want to argue that the costs exceed the benefits" [16, p. 148].

In truth, particle physics can yield practical benefits. The world wide web was a spinoff invention from the planning phase of the LHC. And medical proton therapy is a spinoff of particle accelerator technology.[31] But these sorts of practical dividends are rare and apparently idiosyncratic. Thus, they are not very helpful as a justification for particle-physics experiments.

So, if proponents of particle physics experiments do not attempt to quantify benefits, how do they argue in favor of their expensive experiments? They provide qualitative statements. Often these statements are emotionally charged. For instance, physicist Stephen Hawking characterized the LHC as "vital if the human race is not to stultify and eventually die out".[32]

---

[30] See, e.g., Mangano, supra note 29 (beginning at 41 min.).

[31] Jeremy N.A. Matthews, Accelerators Shrink to Meet Growing Demand for Proton Therapy, Physics Today, March 2009, at 22, available at http://ptonline.aip.org/journals/doc/PHTOAD-ft/vol_62/iss_3/22_1.shtml.

A recurrent theme in the non-quantified argument for particle colliders is to make a special claim of importance for particle physics over other scientific fields. An example is what Nobel laureate Steven Weinberg wrote in his book *Dreams of a Final Theory*: "The reason we give the impression that we think that elementary particle physics is more fundamental than other branches of physics is because it is, I do not know how to defend the amounts being spent on particle physics without being frank about this" [21, p. 55].

Anthropologist Sharon Traweek puts it this way [19, p. 2–3]:

> The physicists' calling is awesome: memoirs and biographies often present this corps d'elite as unique, Promethean heroes of the search for truth. …The extraordinary scale and costliness of much physics research if anything reinforces its cultural value. The great accelerators, for example, are like medieval cathedrals: free from the constraints of cost-benefit analysis.

Thus, particle physics experimentation is a project for which no practical benefit is anticipated or sought. The upside is philosophic, and if the goal is "unmasking the cosmos,"[33] it is indubitably sublime. Once the benefit has been taken out of the realm of numbers, quantitative cost-benefit analysis is rendered moot.

### 6.4.2 Moving Away from the "Probability Mode"

Scientists advocating large-scale leading-edge science experiments can also repel criticism by de-quantifying the discussion of risk on the other end of the cost-benefit formula—by refusing to use numerical values to discuss the chance of disaster.

Following its experience with questions about RHIC safety, physicists learned to keep quantified probabilities out of the LHC safety debate. By doing so, physicists could avoid uncomfortable questions of what constitutes an acceptable level of risk of planetary destruction. Instead, the disaster question was cast as binary: either the experiment entails a risk, or it does not.

A window into how physicists plan the public-relations side of risk assessment comes from the video of a presentation given by CERN theorist John Ellis to his colleagues at the laboratory.[34] In the discussion following the presentation, an audience member said to Ellis, "I've noticed that, very wisely, you haven't pronounced the word 'probability'."[35]

---

[32] Jon Swaine, Stephen Hawking: Large Hadron Collider vital for humanity, The Telegraph, September 9, 2008, http://www.telegraph.co.uk/news/2710348/Stephen-Hawking-Large-Hadron-Collider-vital-for-humanity.html.

[33] See Brian Greene, The Fabric of the Cosmos 22 (Alfred Knopf 2004) ("[A]s we've continued to unmask the cosmos, we've gained the intimacy that comes only from closing in on the clarity of truth. The explorations have far to go, but to many it feels as though our species is finally reaching childhood's end.").

[34] See video recording, supra note 27.

[35] See id. (beginning at 64 min.).

"Absolutely," Ellis said.

The audience member noted that when probability comes into the debate, critics can easily make an argument that the LHC is not worthwhile. That is because, no matter how small the likelihood of destruction, since the harm is so enormous, operating the LHC may be painted as a poor choice.

The lopsided balancing of costs and benefits is not the only problem created for experiment proponents by speaking in terms of probability. There is also the problem that when scientists provide a quantitative assessment in the form of a probability bound—that is, a worst-case limit on how likely a disaster could be—this ceiling on risk may be consumed by the public as if it were an estimate of the actual probability of disaster. To take a hypothetical example, a probability bound of one in a billion does not mean that an event is likely to occur once in a billion trials. It means that the risk is not more than one in a billion—even though in reality it might be far less.

Probability bounds are not actual probabilities, and thus it is fallacious to equate the two. But it is not necessarily fallacious to take probability bounds as reasonable stand-ins for probabilities. In thinking through the ultimate question of whether an experiment should be given the green light, it may be a sensible analytical step to think in terms of the worst-case scenario—which the probability bound represents. Doing so seems particularly apropos when the endeavor under consideration has only a nebulous, philosophic benefit. Yet sensible or not, such worst-case-scenario reasoning works solely against the interests of experimenters. Thus, the disincentive to produce quantified probability bounds remains.

After the flap over the RHIC's odds, CERN steered clear of assigning probabilities for LHC/black-hole risk, instead issuing unquantified statements disclaiming all risk. The Giddings and Mangano paper provided no quantified odds; rather, it made the qualitative statement that there was "no risk of any significance whatsoever" [7]. Note that this statement allows that some possibility of danger exists, but whatever quantitative extent that risk might have is veiled behind the value judgment that the risk is insignificant.

One problem with such qualitative statements is that they are susceptible to recharacterization. For instance, the LSAG report on the black-hole issue, while relying on the Giddings and Mangano paper to provide its rationale, framed its conclusion with a rosier qualitative statement, saying that LHC-generated black holes "present no conceivable danger".[36] Upon receipt of the LSAG report, CERN's permanently constituted Scientific Policy Committee went even further, saying that the LSAG report provided "proof" that the LHC was safe.[37]

There can also be coordination of rhetoric among particle physicists in an effort to help shape public debate. As reported in a 2007 New Yorker magazine article, CERN's chief science officer Jos Engelen explained that, when it comes to LHC disaster scenarios, CERN officials are instructed "not to say that the probability is very small but that the probability is *zero*".[38]

---

[36] LSAG Version 2, at p. 1.

[37] Peter Braun-Munzinger, Matteo Cavalli-Sforza, Gerard 't Hooft, Bryan Webber, & Fabi Zwirner ("CERN Scientific Policy Committee"), CERN, SPC Report on LSAG Documents 1 (2008), http://indico.cern.ch/getFile.py/access?contribId=20\&resId=0\&materialId=0\&confId=35065.

Another example along these lines comes from Ellis's account of his interactions with Cambridge University's Martin Rees—Britain's Astronomer Royal and a CERN outsider. When Rees stated that the risk of the LHC causing disaster was no more than one in 50 million,[39] Ellis reached out to him.

"I … extracted from him a statement that he'd never done an estimate, and he doesn't believe there's any risk. So he's also gone over to the not-talking-about-probability mode," Ellis said. "But I'm keeping his statement in my mail until such time as this issue raises its head".[40]

Ellis said in 2008 that, since the LSAG report came out, he had seen no discussion of risk in the probability mode.[41]

### 6.4.3 Constructing the Quantum Straw Man

Another way in which LHC proponents re-framed the debate to advantage was to make use of a particular kind of uncertainty—quantum uncertainty—as a way to paint the black-hole question as silly.

Nima Arkani-Hamed, a particle physicist at Princeton, proffered the argument in perhaps its most colorful and memorable form to The New York Times when he explained that there was a minuscule probability "the Large Hadron Collider might make dragons that might eat us up".[42]

Engelen offered an expanded version of the argument to the New Yorker: "In quantum mechanics, there is a probability that this pen will fall through the table", Engelen said. "All of a sudden, it will be on the floor. Because it can behave as a wave, it can go through; we call that the 'tunnel effect.' If you calculate the probability that this happens, it is not identical to zero. It is a very small probability. But it never happens. I've never seen it happen. You have never seen it happen. But to the general public you make a casual remark, 'It is not identical to zero, it is very small,' and …"[43] The reporter indicates Engelen then shrugged.

---

[38] See Elizabeth Kolbert, CRASH COURSE Can a seventeen-mile-long collider unlock the universe? The New Yorker (May 14, 2007), http://www.newyorker.com/reporting/2007/05/14/070514fa_fact_kolbert.

[39] Video recording, supra note 27 (beginning at 65 min.).

[40] Id. (beginning at 66 min.).

[41] Id. (beginning at 64 min.).

[42] See Dennis Overbye, Asking a Judge to Save the World, and Maybe a Whole Lot More, New York Times, Mar. 29, 2008, at A1, available at http://www.nytimes.com/2008/03/29/science/29collider.html.

[43] Kolbert, supra note 26.

80	E.E. JohnsonThe quantum-chance-of-anything argument was embraced by the press and the blogosphere, constituting a clear public-relations victory.[44] Yet it is fallacious. The quantum dragon is a straw man. No critic of the LHC argued that the collider should be shut down because of a generic quantum-mechanics-type chance of producing a planet-eating black hole. The argument that the LHC might destroy the planet is that, under current theory, the LHC, owing to its novel characteristics, might produce black holes, and there is no good way to rule out that such a black hole would destroy the planet.

What is especially slippery about the quantum-dragons argument is that it is clothed in the language of particle physics. By incanting "quantum mechanics", the argument seems to claim some particular relevance to particle accelerators. But that is not the case. Since there is a quantum-mechanical chance of just about anything happening, the quantum-dragons-type of straw man can be applied to any debate. For instance, a person arguing for the safety of tobacco could make the same argument in response to the allegation that cigarettes cause cancer. Such a person could correctly point out that there is, quantum mechanically speaking, a chance that cigarettes will turn into tiny sticks of dynamite and explode. The argument is just as irrelevant to a debate about tobacco as it is to the LHC.

## 6.5  Implications for Courts

So far, I have argued that uncertainty can make science-experiment risk inapt for traditional risk assessment and that traditional modes of thinking about risk allow adroit proponents of scientific experiments to reframe questions of risk in a way that is favorable for the experimenters. Now I will discuss what this means for courts in the context of a legal challenge to an experimental program based on the allegation that it is unduly risky.

At the outset, it should be observed that debates about the safety of science that take place in the media are likely to have an effect on litigation outcomes. Of course, courts are ideally meant to be instruments of the rule of law, and public opinion has no direct relevance to deciding issues such as whether a court should use its equitable powers to halt a scientific experiment. Yet it would be naïve to say that public opinion does not carry a great deal of weight inside a courtroom. As Chief Justice William H. Rehnquist of the United States Supreme Court observed, "Judges, so long as they are relatively normal human beings, can no more escape being influenced by public opinion in the long run than can people working at other jobs" [18]. Thus, a victory won through the media in the court of public opinion becomes a substantial hurdle to obtaining an injunction, should one be merited.

---

[44]See, e.g., Sharon Weinberger, Collider May End World! (Or Spit out Man Eating Dragons), Wired.com Danger Room, April 16, 2008, http://blog.wired.com/defense/2008/04/collider-may-en.html; Dennis Overbye, Gauging a Collider's Odds of Creating a Black Hole, New York Times, April 15, 2008, http://www.nytimes.com/2008/04/15/science/15risk.html. ("Besides, the random nature of quantum physics means that there is always a minuscule, but nonzero, chance of anything occurring, including that the new collider could spit out man-eating dragons."). Bloggers and blog commenters citing this argument are too numerous to cite.

Additionally, the dynamics of the debate outside the litigation context are likely to mirror debate that would take place in court. The same ways of framing questions of risk and asserting uncertainty that are persuasive to the public are likely to be persuasive to a judge as well.

Bearing this in mind, there are ways to make questions of risk from large-scale leading-edge science experiments more meaningfully susceptible to judicial review. Here I will propose two. The first is a primary, general strategy: evaluating uncertain risks in qualitative terms. Then I discuss a secondary, adjunct strategy: testing the favorable opinions of science-experiment proponents against their opinions outside the context of the controverted experiment.

### 6.5.1 Evaluating Uncertain Risks in Qualitative Terms

Although courts have often looked at risk in quantitative terms,[45] in the case of science experiments where uncertainty dominates, courts should evaluate risk in a qualitative sense, conducting a kind of meta-analysis that gets above the level affected by uncertainty.[46]

This approach involves framing the uncertainty in human terms—in particular, looking to the context of scientists' assessment of their own experiment as safe. The assumption here is that experts can be influenced in their judgments on risk by their cultural orientation [13]. That is to say, experts' reasoning is vulnerable to biasing influences [13, p. 1081–1082 and 1093–1094]. As legal scholar Dan Kahan and co-authors wrote, "Like members of the general public, experts are inclined to form attitudes toward risk that best express their cultural vision" [13, p. 1094].

To form a qualitative assessment of the extent to which cultural vision could affect safety assessments about the risk of experiments, courts should look at the scientists' organizational culture, community norms, group politics, and power dynamics. These human elements may provide reasons to be confident in, or skeptical of, the scientists' judgments about the risk of their own experimental programs.

As a complement to looking at the cultural context, it will likely be of use to look at the safety argument itself. This is because certain aspects of the scientists' argument may be relevant in evaluating susceptibility to bias and cultural filtering. Simple arguments for safety will be more resistant to such biases and filters. No amount of bias or cultural filtering would, for instance, cause someone to believe 2 + 2 = 5. But the more complex the chain of reasoning involved, the more opportunity there is for judgment calls regarding what assumptions to make and what data to reference. That, in turn, means there is more room for bias and cultural influence to determine scientists' own judgment about the riskiness of their experiments.

---

[45]See, e.g., Hawai'i County Green Party, 980 F. Supp. at 1168–69 (using quantified costs and probabilities in denying an application for injunction to halt launch of the Cassini space probe carrying 32.8 kg of plutonium as a power source, which plaintiffs alleged caused a cancer risk in the event of a launch accident or navigation error).

[46]What I am suggesting here is an adaptation of an approach I suggested previously for the related problem of courts dealing with scientific questions of extraordinary complexity that are opaque to laypersons [11, p. 883–907].

One can imagine readily applying this sort of analysis to the LHC/black-holes case. Some factors counseling skepticism of the risk assessment offered by CERN include the involvement of people with career stakes in the outcome of the assessment, the complexity of the safety case, and the fact that the assumptions and models used in the risk analysis require the exercise of discretionary judgment.

It may seem to be a step backward in dealing with risk to intentionally embrace qualitative reasoning in lieu of quantitative analysis. But we really have no choice. As discussed above, quantitative analysis can be rendered mostly meaningless on account of its uncertain assumptions and propensity for error. Under these circumstances, quantitative sophistication has an illusory allure, while qualitative analysis, if less prepossessing at first glance, offers more in the way of integrity.

### 6.5.2 Testing the Opinions of Science-Experiment Proponents by Analogy to Their Opinions Outside the Context of the Controverted Experiment

Beyond generally moving to a qualitative mode to evaluate risk, it may be probative to see to what extent scientists' assessments of the safety of their own experiments are consistent with their assessments in areas outside their own sphere of self-interest.

A simple example comes from Stephen Hawking, who opined that the LHC is "absolutely safe". Yet speaking outside of the context of LHC safety, he wrote that it is plausible that current physics theory may someday be regarded "as ridiculous as a tower of turtles" [9]. Given Hawking's understanding that today's physics theory may not endure, there may be some reason to doubt his ironclad conclusions about LHC risk.

A better example, with much richer detail, is the writing that particle physicist Lisa Randall has done about the LHC/black-hole question and about risk in the financial sector.

Randall is a proponent of the LHC, saying that the risk of the accelerator generating dangerous black holes is "essentially nonexistent" [17, p. 179]. She does admit that some uncertainties are involved when it comes to the question of accelerator-created black holes.[47] Yet Randall's bottom-line assessment is that uncertainties in the LHC/black-hole risk question can be safely ignored [17, p. 186–187]:

> Luckily for our search for understanding, we are extremely certain that the probability of producing dangerous black holes is minuscule. We don't know the precise numerical probability for a catastrophic outcome, but we don't need to because it's so negligible. Any event that won't happen even once in the lifetime of the universe can be safely ignored.

It is interesting to compare Randall's thinking about the LHC/black-hole question to how she views risk in the financial system. Randall advocates additional regulation for the financial system. Why? Randall urges that attention should be paid even to unlikely outcomes if the harm would be very large. With insight that could easily be applied to the black-hole case, she writes [17, p. 186]:

> The financial crisis happened because of events that were outside the range of what the experts had taken into account. …Virtually no one paid attention to the 'unlikely' events that precipitated the crisis. Risks that might otherwise have been apparent therefore never came up for consideration. But even unlikely events need to be considered when they can have significant enough impact. Any risk assessment is plagued by the difficulty of evaluating the risk that the underlying assumptions are incorrect. Without such estimates, any estimate becomes subject to intrinsic prejudices. On top of the calculational problems and hidden prejudices buried in these underlying assumptions, many practical policy decisions involve unknown unknowns—factors that can't be or haven't been anticipated. Sometimes we simply can't foresee the precise unlikely event that will cause trouble. This can make any prediction attempts—that will inevitably fail to factor in these unknowns—completely moot.

Similarly, Randall explains why we cannot rely on experts in the realm of finance and economics [17, p. 195]:

> After all, 'experts' told us that derivatives were a way of minimizing risk, not creating potential crises. 'Expert' economists told us that deregulation was essential to the competitiveness of American business, not to the potential downfall of the American economy…Clearly experts can be shortsighted. And experts have conflicts of interest.

Yet Randall exempts particle physicists from such fallibility. This is despite the fact that physicists can be shortsighted. Recall that physicists in 1999 prematurely declared that present-day accelerators would not have nearly enough energy to create black holes, only to be contradicted by evolving theory a couple years later. And physicists, of course, can have conflicts of interest. Recall that CERN's safety assessment was done by CERN affiliated individuals.

Notably, Randall seems to appreciate the tension in her views. She writes, "Yet despite my confidence that it was okay to rely on experts when evaluating potential risks from the LHC, I recognize the potential limitations of this strategy and don't quite know how to address them" [17, p. 195].

---

[47] See [17] at 172 ("No one really knows how to solve systems in which both quantum mechanics and gravity play an essential role. String theory is physicists' best attempt, but we don't yet understand all its implications. This means that in principle there could be a loophole. Extremely tiny black holes, which we will understand only with a theory of quantum gravity, are unlikely to behave the same way as the big black holes we derive using classical gravity. Perhaps such very tiny black holes don't decay at the rates we expect. Even this isn't a serious loophole however.") (internal paragraph break omitted).

## 6.6 Conclusion

We must accept the fact that it is in our nature as humans to push the frontiers of our knowledge. And we must accept as well that we can be mistake-prone and shortsighted in doing so. If the LHC does not present a real danger to humanity, another great science experiment soon might. Acknowledging our inability to tame uncertainty is the first step in being prudent and rational with such issues. Developing intelligent means of navigating uncertainty is the next step. And it's in that vein that I have sought to make a contribution with this chapter.

While the upshot of my analysis is that the courts can, indeed, do a good job in handling uncertainty in the science-experiment-risk context, a word should be said in closing about whether courts should perform this function. Without engaging in a general defense of the rule of law, I will note that when it comes to uncertainty and risk, this book points out at least two relevant desiderata. In Chap. 7, Jordan Sand takes away from the history of Edo the lesson that giving people control over their fates provides a way to live with uncertainty. And in the book's conclusion, Corinne Bieder emphasizes the key role of trust.

The courts are well-positioned to provide both trust and a meaningful sense of control. The courts supply an avenue to trust through their role of gathering and impartially weighing evidence. And the openness of the courts to hearing complaints of affected parties can provide people everywhere with a sense of control over their own destinies. In the overall analysis, it does not appear that the existence of uncertainty militates against courts engaging in questions of uncertain risk. Instead, thoughtful reflection seems to show that judicial resolution is particularly appropriate.

By proposing tools for generalist courts to deal with risk questions such as those explored here, I do not mean to imply that engaging with such issues will be easy. To the contrary, there seems little doubt that such cases will be extremely challenging. But if we can achieve so much success in reaching new heights of knowledge about the world around us, certainly we can advance our ability to make good decisions about how to do so safely.